\documentclass[doublecol]{epl2} 
\usepackage{graphicx,amsfonts,amssymb,amsmath,hyperref,enumerate}
\usepackage[]{algorithm2e}

\newif\ifhyper
\hypertrue
\ifhyper
\hypersetup{
   citecolor = {green},
   colorlinks = {true}, % false
   urlcolor = {blue} % magenta
}
\fi

\newcommand{\beq}{\begin{equation}}
\newcommand{\eeq}{\end{equation}}
\newcommand{\beqa}{\begin{eqnarray}}
\newcommand{\eeqa}{\end{eqnarray}}
\newcommand{\ket} [1] {\vert #1 \rangle}
\newcommand{\bra} [1] {\langle #1 \vert}

\def\bra#1{\langle#1\vert}
\def\ket#1{\vert#1\rangle}

\def\Longarrow{\protect\@lra}
\def\@lra{\relbar\joinrel\relbar\joinrel\relbar\joinrel%
          \relbar\joinrel\rightarrow}

\title{Entanglement Continuous Unitary Transformations}

\author{Serkan Sahin \inst{1} \and Kai Phillip Schmidt \inst{2} \and Rom\'an Or\'us \inst{1}}
\institute{
\inst{1} Institute of Physics, Johannes Gutenberg University, 55099 Mainz, Germany \\Ê
\inst{2} Institut f\"ur Theoretische Physik, Universit\"at Erlangen-N\"urnberg, Staudtstra\ss e 7, 91058 Erlangen, Germany
}


\abstract{
Continuous unitary transformations are a powerful tool to extract valuable information out of quantum many-body Hamiltonians, in which the so-called flow equation transforms the Hamiltonian to a diagonal or block-diagonal form in second quantization. Yet, one of their main challenges is how to approximate the infinitely-many coupled differential equations that are produced throughout this flow. Here we show that tensor networks offer a natural and non-perturbative truncation scheme in terms of entanglement. The corresponding scheme is called ``entanglement-CUT" or eCUT. It can be used to extract the low-energy physics of quantum many-body Hamiltonians, including quasiparticle energy gaps. We provide the general idea behind eCUT and explain its implementation for finite 1d systems using the formalism of matrix product operators. We also present proof-of-principle results for the spin-1/2 1d quantum Ising model and the 3-state quantum Potts model in a transverse field. Entanglement-CUTs can also be generalized to higher dimensions and to the thermodynamic limit. 
}

\begin{document}

\maketitle

\emph{Introduction.-} The study of strongly correlated systems entails some of the most important challenges in modern physics. As an example, it is still not fully clear whether the fermionic Hubbard model captures high-$T_c$ superconductivity in cuprates \cite{hubbard, highTc}, or whether some frustrated Heisenberg antiferromagnets can stabilize a quantum spin liquid ground state or not \cite{KHAF}. It is mostly because of challenges like these that many numerical simulation methods were proposed over the years. Still, each one of these methods comes with its own limitations. For instance, exact diagonalization is restricted to systems of relatively small size, and quantum Monte Carlo simulations are hampered by the infamous sign-problem \cite{sign}.  

In this context, continuous unitary transformations (CUTs) have become one poweful tool to assess quantum many-body systems \cite{CUT}. In CUT-based methods the Hamiltonian is continuously transformed by some unitary operator that depends on a continuous parameter $\ell$. This unitary is constructed in such a way that, when $\ell \rightarrow \infty$, the effective Hamiltonian becomes diagonal or block-diagonal in second quantization allowing to extract the important physical properties of complex quantum many-body systems more easily. Nowadays, there are different ways of constructing unitary operators satisfying this property \cite{CUT,generators, Knetter2000, deo}. One of the limitations of this technique, however, is that the evolution of the matrix elements of the Hamiltonian is described by a system of coupled differential equations, and the number of these equations increases with the flow in $\ell$, quickly becoming intractable. As a consequence, several truncation schemes have been developed over the years which are for example based on perturbation theory \cite{Knetter2000, pcut}, real-space approaches using non-perturbative linked-cluster expansions \cite{gcut} or the spatial extension of operators \cite{sCUT}, or in momentum space \cite{Heidbrink2002} using the scaling dimension of operators \cite{scalingdim}.   

Another family of methods that has become quite popular in recent years is based on tensor networks (TNs) \cite{TN}. Here the wavefunction of the quantum many-body system is represented by a set of interconnected tensors that make explicit the natural pattern of entanglement and correlations in the system. Famous examples of TN methods include the density matrix renormalization group (DMRG) \cite{DMRG} and the time-evolving block decimation (TEBD) \cite{tebd} for 1d systems, as well as algorithms based on projected entangled pair states (PEPS) \cite{PEPS} for 2d systems, and the multi-scale entanglement renormalization ansatz (MERA) \cite{MERA} for critical systems. Using these methods it is possible to approximate with good accuracy low-energy as well as time-dependent properties. TN algorithms can be applied to finite- and infinite-size systems \cite{iTEBD, iPEPS, iDMRG}, as well as to fermions \cite{fermions}. Their only limitation is the amount and structure of entanglement in the wavefunction to be described. 

Seeing the above in perspective, one might wonder whether the CUT can be formulated in terms of a TN structure that can be used efficiently to truncate the flow non-perturbatively in terms of entanglement. The main point of this paper is that this is indeed the case. More precisely, we show that TNs offer a very natural approximation of the CUT equations, by truncating in the operator-entanglement content throughout the flow. We call this new scheme \emph{entanglement-CUT} (eCUT). This idea can in principle be applied to any dimensionality and system size (including infinite) as long as a faithful TN description of the flow can be set up. 

For the sake of concreteness, here we provide an explicit prescription for eCUT applied to finite 1d systems, using the formalism of matrix product operators (MPOs) \cite{TN} and we give proof-of-principle results for the spin-1/2 1d quantum Ising chain and 3-state quantum Potts model in a transverse field and open boundary conditions up to $50$ sites. We show that this allows to determine with good accuracy the ground-state energy as well as the 1st excited state -- amounting usually to the one-quasiparticle (1QP) energy gap --. Moreover, we implement a \emph{ground state generator} for the unitary transformation, which we shall also discuss as a particular case of a \emph{block generator}. 

\emph{CUT and the flow equation.-} Here we review briefly some basic properties of CUT that will be needed later (the interested reader is referred to, e.g., Ref.\cite{CUT} and references therein for more information). Let us start by considering an Hamiltonian $\mathcal{H}$ describing a quantum many-body system that we wish to diagonalize. It is well-known \cite{CUT} that this can be achieved by a unitary transformation that depends on a continuous parameter $\ell$, i.e., 
\beq
\mathcal{H}(\ell) = U(\ell) \mathcal{H} U^{\dagger}(\ell) \ . 
\label{flow}
\eeq
In this equation, $U(\ell = 0) = U(\ell \rightarrow \infty) = {\mathbb I}$. Ideally, the unitary transformation $U(\ell)$ yields a diagonal effective Hamiltonian $\mathcal{H}(\ell \rightarrow \infty)$. In practice, though, one targets a block-diagonal Hamiltonian. The flowing Hamiltonian Eq.~\eqref{flow} can be cast in terms of a generator $\eta(\ell)$, such that 
\beq
\partial_\ell \mathcal{H}(\ell) = [ \eta(\ell) , \mathcal{H}(\ell) ] \ .
\eeq
The above is called the \emph{flow equation}. This generator can be written in terms of the unitary transformation throughout the flow as 
\beq
\eta(\ell) =  U^{\dagger}(\ell) \partial_\ell U(\ell) \ .
\eeq
In the same way, the unitary operator can be written in terms of the generator as
\beq
U(\ell) = \mathcal{T}_\ell \exp{\left(\int_0^\ell d\ell' \eta(\ell')\right)}
\label{int}
\eeq
where $\mathcal{T}_\ell$ means ``$\ell$-ordering". It is easy to check that the generator $\eta(\ell)$ is antihermitian. This generator must also be chosen such that the flow drives the Hamiltonian towards some block-diagonal form, from which the low-energy properties should be easy to extract. In this paper we choose to work with what we call the \emph{block-generator}, which we define through its matrix elements as 
\beq
\eta_{ij}^{\rm b}(\ell) \equiv \sum_{n=0}^r \left( \mathcal{H}_{in}(\ell) \delta_{nj} - \delta_{in}\mathcal{H}_{nj}(\ell) \right)\,. 
\label{bgen}
\eeq
In the above equation, $\mathcal{H}_{ij}$ are the matrix elements of the Hamiltonian in some relevant quasiparticle basis \emph{which is decided beforehand}. In practice, the goodness of the choice of basis is checked a posteriori by the goodness of the results (intuitively one chooses a quasiparticle picture that is known to be exact in some limiting regime, e.g., low or large magnetic fields).  Also, $r$ is the size of the (low-energy) block being targeted, e.g., for $r=0$ one isolates ``just'' the ground state, while for $r=1$ one separates the ground state and the first excitation from all other states. For $r=0$, this generator corresponds to the so-called ``variational ground-state generator" \cite{deo}, which decouples the ground state. For arbitrary $r$, the block generator $\eta^{\rm b}$ decouples the $r$-dimensional low-energy subspace, see Appendix.

\begin{figure}
\includegraphics[width=7.5cm,angle=0]{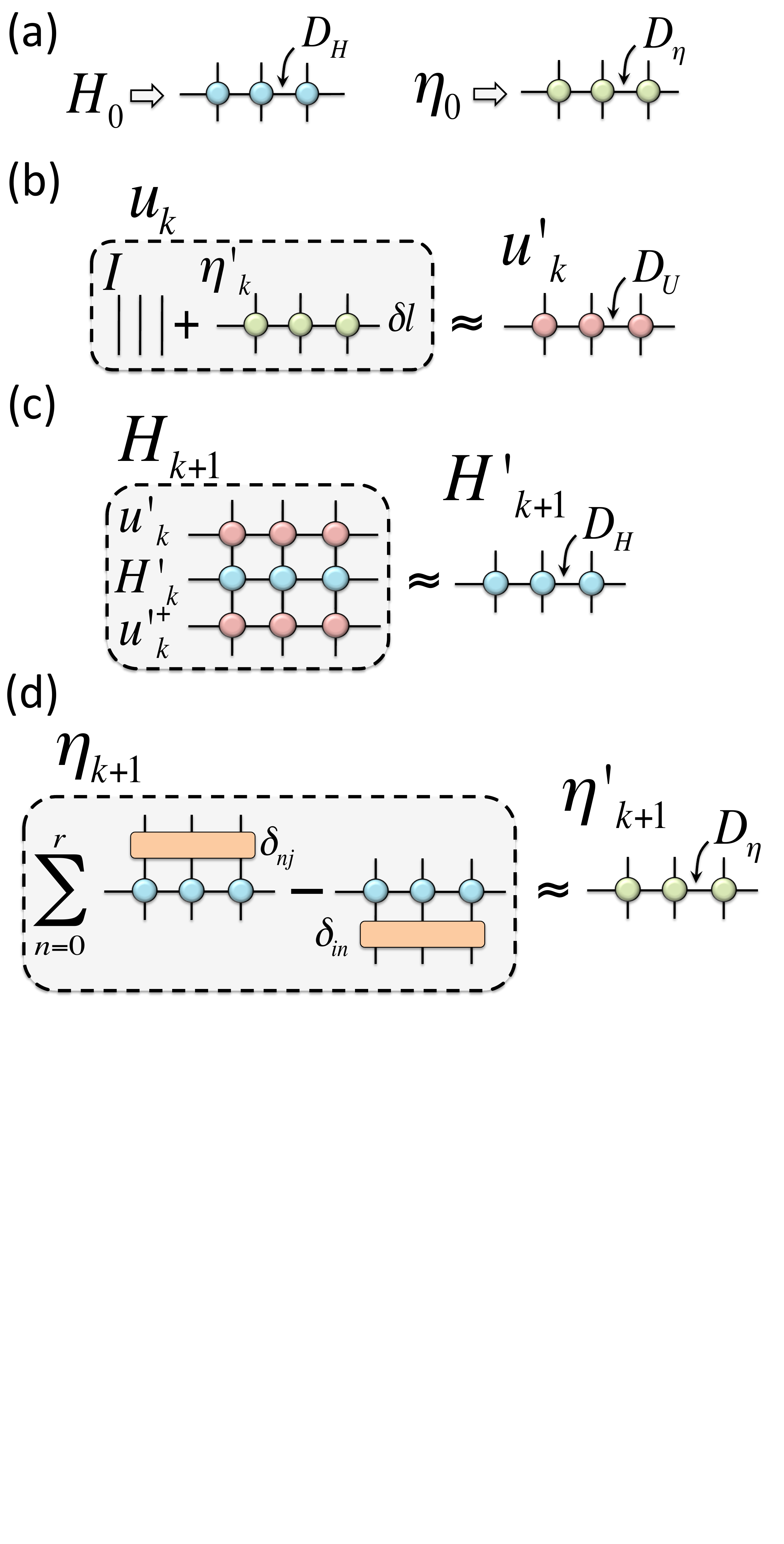}
\caption{[Color online] (a) $\mathcal{H}_0$ and $\eta_0$ are represented as MPOs with bond dimensions $D_{\mathcal{H}}$ and $D_\eta$ respectively. (b) The first-order Taylor expansion of the unitary $u_k$ is approximated by an MPO with bond dimension $D_U$. (c) The evolved Hamiltonian $\mathcal{H}_{k+1}$ is approximated by an MPO with bond dimension $D_{\mathcal{H}}$. (d) The block-generator $\eta_{k+1}$ is approximated by an MPO with bond dimension $D_\eta$, where, e.g., $\delta_{nj}$ is a Dirac delta in the basis of the block being targeted. For a code-implementation example in the ground-state case, see Algorithm \ref{alg}.}
\label{fig1}
\end{figure}

\emph{MPOs and operator-entanglement.-} Operators with an inner 1d structure admit a TN representation in terms of so-called MPOs, see Fig.~(\ref{fig1}) for a diagrammatic representation. These are the natural operators acting on the matrix product states (MPS), which are a family of wavefunctions suitable to account for the properties of 1d gapped Hamiltonians with local interactions \cite{mps}. At every site of an MPO there is a tensor with two \emph{physical indices} of size $d$ each, accounting for the $d$ degrees of freedom of the local Hilbert space at every site (e.g., $d=2$ for spin-1/2). Tensors in the MPO are also connected by \emph{bond indices} with \emph{bond dimensions} $D^{[i]}$, which in principle can depend on the site $i$. These bond dimensions measure the \emph{operator-entanglement} content of the MPO \footnote{More specifically, the singular values $\{\lambda^{[i]}_\alpha \}$ for the bipartition $[ 1, 2, \ldots, i : i+1, i+2, \ldots , N]$, if normalized such that $\sum_{\alpha = 1}^{D^{[i]}}  (\lambda^{[i]}_\alpha)^2 = 1$, satisfy then the constraints $-\sum_{\alpha = 1}^{D^{[i]}}  (\lambda^{[i]}_\alpha)^2 \log (\lambda^{[i]}_\alpha)^2 \leq \log D^{[i]}$, which are nothing but the MPO ``operator-version" of the constraints for the entanglement entropies of an MPS \cite{TN}.}. In practice, we simply call ``bond dimension of an MPO" the parameter $D$ such that $D^{[i]} \le D ~ \forall i$.

\emph{eCUT.-} The main idea of eCUT is to represent the operators for $\mathcal{H}(\ell)$, $U(\ell)$ and $\eta(\ell)$ in Eqs.~(\ref{flow}-\ref{bgen}) by TNs. As the flow in $\ell$ proceeds, the operator-entanglement content needed to represent these operators may grow, and so will do the bond dimensions. Therefore, the bond indices need to be truncated  throughout the flow using standard TN procedures. To achieve this, the flow is discretized in ``flow steps" in $\ell$ (similar to the discretization of time in TEBD \cite{tebd}), and the truncations are done at each discrete step. Generically, such truncations amount to keeping the degrees of freedom in the TN accounting for the majority of the operator-entanglement, which naturally (and indirectly) close the system of coupled differential equations being generated by CUT. 

Let us be more specific with how the algorithm works for a finite 1d system. The core of the method is to consider the $\ell$-ordered integral in Eq.~(\ref{int}) and break it into $M$ smaller steps of size $\delta \ell$, that is, 
\beq
U(\ell) \approx  u_{M-1} u_{M-2} \cdots u_2  u_1 u_0 \ ,
\eeq
with $\ell = M\delta \ell$ and the \emph{unitary transformation at step $k$} being defined as
\beq
u_k \equiv \exp{(\eta_k \delta \ell)} \ .
\eeq
In this aproximation, the error at every step is $O(\delta \ell ^2)$. In the above equation, $\eta_k$ is the \emph{generator at step $k$}, which in turn depends on $\mathcal{H}_k$, the \emph{Hamiltonian at step $k$}, through, e.g., Eq.~(\ref{bgen}). In what follows we show how to approximate $\mathcal{H}_k, \eta_k$ and $u_k$ at every step $k$ using MPOs for a system in 1d. The procedure reads as follows: 

\RestyleAlgo{boxruled}
\begin{algorithm}
$d \leftarrow$ local physical dimension\; 
$I \leftarrow d \times d$ identity\;     
\For{all sites}{     
$A \leftarrow$ local MPO tensor for $H$\;
$D_H \leftarrow D_{Hb} \leftarrow A$ bond dimension; $v \leftarrow 1$\; 
 \If{boundary}{
             $D_{Hb} \leftarrow 1$; {\rm {\bf if}} \emph{left} {\rm {\bf then}} $v \leftarrow -1$\;}
\For{$s=1$  {\rm {\bf to}}  $d$ {\rm {\bf and}}  $t=1$  {\rm {\bf to}}  $d$}{      
            $B_{1:D_H,1:D_{Hb}}^{s,t} \leftarrow A_{1:D_H,1:D_{Hb}}^{s,1}*I^{1,t}$\;
            $B_{D_H+1:2D_H,D_{Hb}+1:2D_{Hb}}^{s,t} \leftarrow v*I^{s,1}*A_{1:D_H,1:D_{Hb}}^{1,t}$\;}
Untruncated generator local MPO tensor $\leftarrow B$}
\caption{Exact ground-state MPO generator, left-hand-side of Fig.(\ref{fig1}.d). Array indices start at $1$.} 
\label{alg}
\end{algorithm}

\underline{1) Initialization:}  at step zero, $k=0$, we write the original Hamiltonian \mbox{$\mathcal{H}_0 \equiv \mathcal{H}(\ell=0)$} as an MPO with bond dimension $D_{\mathcal{H}}$, as shown in Fig.~(\ref{fig1}.a). We also write the original generator $\eta_0 \equiv \eta(\ell=0)$ as another MPO with bond dimension $D_{\eta}$, see Fig.~(\ref{fig1}.a). Usually it is easy to get an exact analytical expression for $\eta_0$, but if this were unknown, one could then even approximate its MPO representation numerically by using Eq.~(\ref{bgen}) and the MPO for $\mathcal{H}_0$. Set also $\mathcal{H}'_0 = \mathcal{H}_0$ and $\eta'_0 = \eta_0$. In what follows, operators with prime will correspond to truncated MPO approximations to operators without prime.  

\underline{2) Main loop:} for $k=0, 1, \ldots, M-1$, iterate the following steps: 

{\emph{(i) Approximate $u_k$:} evaluate $u_k$ using a Taylor expansion, e.g., 
\beq
u_k = \exp{(\eta_k \delta \ell)} = {\mathbb I} +  \eta_k \delta \ell + O(\delta \ell^2) \sim {\mathbb I} +  \eta'_k \delta \ell  \sim u'_kÊ, 
\label{taylor}
\eeq
where we stopped at first order, but higher orders can also be considered. This equation is the addition of two MPOs: one for ${\mathbb I}$, and one for $\eta'_k \delta \ell $, see Fig.~(\ref{fig1}.b). This, in turn, can be approximated by an MPO of bond dimension $D_U$ for an operator $u'_k$, which is our approximation to $u_k$. 

{\emph{(ii) Approximate $\mathcal{H}_{k+1}$:} approximately evolve the Hamiltonian by a step $\delta \ell$, namely 
\beq
\mathcal{H}_{k+1} = u_k \mathcal{H}_k u_k^{\dagger} \sim u'_k \mathcal{H}'_k u_k^{\prime \dagger}  \sim \mathcal{H}'_{k+1}\ .
\eeq
The TN diagrams representing this equation are in Fig.~(\ref{fig1}.c). The approximation on the right hand side means that the resulting tensor network for $\mathcal{H}_{k+1}$ is approximated by an MPO of bond dimension $D_\mathcal{H}$ for an operator $\mathcal{H}'_{k+1}$. 

{\emph{(iii) Approximate $\eta_{k+1}$:} compute the generator $\eta_{k+1}$ assuming,  e.g., the block generator in Eq.~(\ref{bgen}) (but others should also be possible), 
\beqa
(\eta_{k+1})_{ij} &=& \sum_{n=0}^r \left( (\mathcal{H}_{k+1})_{in}(\ell) \delta_{nj} - \delta_{in}(\mathcal{H}_{k+1})_{nj})(\ell) \right)Ê\nonumber \\Ê
&\sim& \sum_{n=0}^r \left( (\mathcal{H}'_{k+1})_{in}(\ell) \delta_{nj} - \delta_{in}(\mathcal{H}'_{k+1})_{nj})(\ell) \right) \nonumber \\
&\sim& (\eta'_{k+1})_{ij} , 
\eeqa
where the last approximation means that we are approximating the resulting operator by an MPO with bond dimension $D_\eta$ for the generator operator $\eta'_{k+1}$, as shown in Fig.~(\ref{fig1}.d). 

The whole algorithm follows then by iterating the main loop, that is
\beq
u'_k \rightarrow \mathcal{H}'_{k+1} \rightarrow \eta'_{k+1} \rightarrow u'_{k+1} \rightarrow \mathcal{H}'_{k+2} \rightarrow \cdots \rightarrow \mathcal{H}'_{M-1}. 
\eeq
In this way we obtain a series of approximated Hamiltonians $\{ \mathcal{H}'_1, \mathcal{H}'_2, \ldots \mathcal{H}'_{M-1} \}$. For sufficiently large $M$, and sufficiently-accurate MPO approximations, the Hamiltonian $\mathcal{H}'_{M-1}$ will converge to a block-diagonal form. For the case of the generator in Eq.~(\ref{bgen}), the low-energy r-dimensional block will be decoupled from the rest in the chosen quasiparticle basis, from which it is now immediate to extract the ground state energy, as well as (potentially) the low-energy excitations. Other observables can be computed by flowing the corresponding operator in parallel with the Hamiltonian. In practice, approximations and truncations for MPOs are done using standard TN techniques for 1d finite-size systems, see the appendix material for more information. Notice that our method amounts to finding a quantum circuit that block-diagonalizes the Hamiltonian which, when run in reverse starting from appropriate product states, will produce the eigenstates of the original Hamiltonian \mbox{$\mathcal{H} = \mathcal{H}(\ell=0)$.} Therefore, this method is based on an evolution in the Heisenberg picture, i.e., we evolve operators rather than states. This strategy is similar to the one used in some techniques to compute the so-called MERA \cite{flowMERA} and, in the continuum-flow limit, the so-called continuous-MERA \cite{cMERA}. The overall leading computational cost of our algorithm is easily checked to be $O(d^6 (D_\eta^3 + D_{\mathcal{H}}^6 D_U^6))$. 

To get the 1st excited state one can consider different strategies. A possibility is to directly consider the block generator for $r > 0$. This is straightforward, but it involves larger bond dimensions in the algorithm, \emph{because multiple eigenstates are being simultaneously targeted in superposition}. An initial numerical test showed that this approach works well for systems of moderate size, but not too large. To overcome this problem, another possibility is to use an ``inverted spectrum technique", flowing with Hamiltonian $\mathcal{H}_I \equiv (\mathcal{H} - \lambda \mathbb{I})^2$, and scanning $\lambda$ to hit the desired excitation. Moreover, for the case of a translationally-invariant system (e.g, with periodic boundary conditions) the 1st excited state is protected by the momentum quantum number $k$ and could therefore be targeted as the ground state in a given $k$-sector. Our plan is to use this strategy in forthcoming implementations of the method. For the purpose of this paper, however, we choose yet a different option to show proof-of-principle results for the excited state, namely, we use the same strategy as for the ground state but with the ``shifted" Hamiltonian $\mathcal{H}_S \equiv \mathcal{H}  + \Delta_S \ket{\Psi_0}\bra{\Psi_0}$, with $\Delta_S$ an energy shift larger than the 1QP gap, and $\ket{\Psi_0}$ an MPS approximation of the ground state of $\mathcal{H}$ (which can be computed with, e.g., TEBD or DMRG). This approach has several advantages: first, the required bond dimensions are not too large since only one quantum state is targeted. And second, the stability is well controlled as in the case of the ground state. 

Finally,  a good practical criteria to determine the best point in the approximated flow is to choose the minimum of the so-called \emph{reduced off-diagonality (rod)}, ${\rm rod} \equiv \frac{1}{N}\sqrt{\sum_{i \neq 0 } | \mathcal{H}_{i0} |^2}$, where $\mathcal{H}_{ij}$ are the Hamiltonian matrix elements and $0$ is the subspace that we wish to decouple. If truncations were exact, this quantity would tend to be exactly zero, and therefore its global minimum over the flow defines the optimal point at which we read the energy. 

\emph{Results.-} To prove the validity of our method we present here benchmarking calculations for the 1d spin-1/2 quantum Ising modelÊ and the 3-state quantum Potts model in a transverse magnetic field and with open boundary conditions. Their Hamiltonians are given by 
\beqa
\mathcal{H}_{Ising} &=& - \sum_{i=1}^{N-1} \sigma_x^{[i]} \sigma_x^{[i+1]} - h \sum_{i=1}^N \sigma_z^{[i]} \nonumber \\  
\mathcal{H}_{Potts} &=& - \sum_{i=1}^{N-1} \left( M^{[i]} \bar{M}^{[i+1]} + hc \right) - h \sum_{i=1}^N Z^{[i]} , 
\eeqa
with $\sigma_{\alpha}^{[i]}$ the $\alpha$th Pauli matrix at site $i$, $N$ the number of sites, and the Potts matrices are given by
\beq
M = 
\begin{pmatrix}Ê
0 &1 & 0 \\Ê
0 &0 &1 \\Ê
1 &0 &0
\end{pmatrix}, 
\bar{M} = 
\begin{pmatrix}Ê
0 &0 &1 \\Ê
1 &0 &0 \\Ê
0 &1 &0
\end{pmatrix}, 
Z = 
\begin{pmatrix}Ê
2 &0 &0 \\Ê
0 &-1 &0 \\Ê
0 &0 &-1
\end{pmatrix}. 
\eeq
In our simulations, we considered a chain of up to $N=50$ spins (though we could easily consider larger sizes), with flow step $\delta \ell = 0.01$, Taylor expansions up to third order, and MPO bond dimensions up to $30$. The quasiparticle basis are either the eigenbasis of $\sigma_x$ ($X$-basis) or $\sigma_z$ ($Z$-basis) for Ising,Ê and of $Z$ for Potts ($Z$-basis). A priori, the $X$-basis is  expected to produce good results for low magnetic fields, whereas the $Z$-basis should provide better results at large fields. We run the flow for up to $1000$ steps, and target the ground state of the system and its 1st excitation using the method mentioned above. 

\begin{figure}
\includegraphics[width=8.5cm,angle=0]{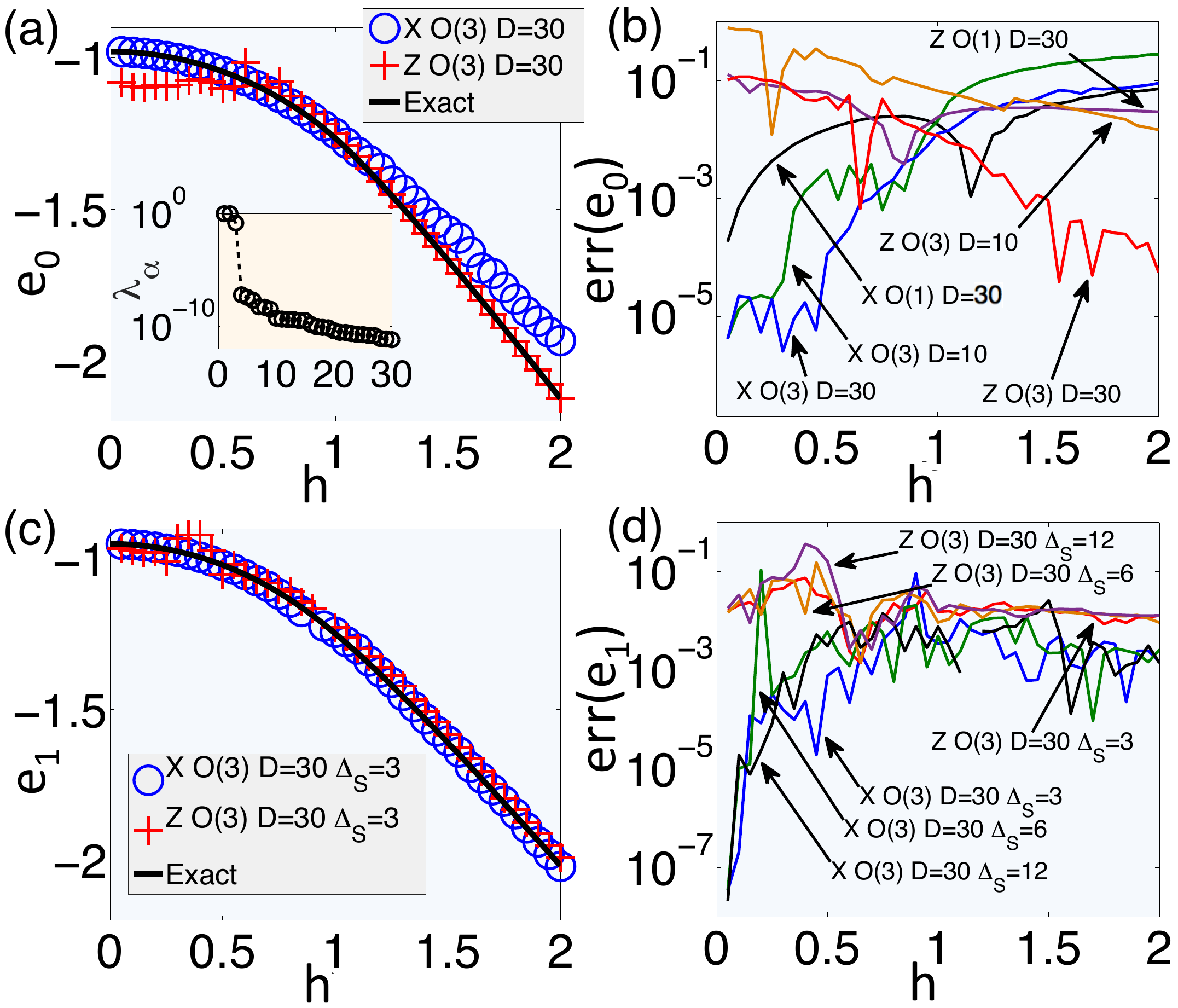}
\caption{[Color online] Quantum Ising model: (a) ground state energy per site $e_0$ as a function of the magnetic field for $N=50$, all MPO bond dimensions upper-bounded by $D=30$, for simulations in the $X$ and $Z$ basis and $O(3)$ Taylor order. The critical point in the thermodynamic limit is at $h_{\rm c}=1$. The inset shows a typical MPO singular value spectrum at some intermediate point in the flow for $h=0.95$, which decays exponentially fast. (b) Relative error in $e_0$ for $N=50$ and several simulations with the $X$ and $Z$ basis, $O(1)$ and $O(3)$ Taylor orders, and $D=\{10, 30\}$ upper bounds in the MPO bond dimensions. (c) 1st excited state energy per site $e_1$ as a function of the magnetic field for $N=20$, all MPO bond dimensions upper-bounded by $D=30$ and energy shift $\Delta_S = 3$, for simulations with the $X$ and $Z$ basis, and $O(3)$ Taylor order. (d) Relative error in $e_1$ for $N=20$ and several simulations with the $X$ and $Z$ basis, $O(3)$ Taylor order, $\Delta_S =\{3, 6, 12\}$, and $D=30$ upper bound in the MPO bond dimensions.}
\label{fig2}
\end{figure}

For quantum Ising, our results for the ground state and $N=50$ are shown in Fig.~(\ref{fig2})(a,b). We can see how the $X$-basis works better for low fields and the $Z$-basis for large fields, as expected. The error decreases rapidly with the Taylor order  (seemingly exponentially fast), as well as with the bond dimension. Also, we observe that the spectrums of singular values in the MPO decompositions decay exponentially fast (even close to criticality), which validates the precision of the truncation. Moreover, in Fig.~(\ref{fig2})(c,d) we show proof-of-principle results for the 1st excited state and $N=20$, using $\Delta_S = \{ 3, 6, 12 \}$ and $\ket{\Psi_0}$ previously computed with TEBD. Surprisingly, in this case the $X$-basis produces already reasonably good results for all field values, even better than the $Z$-basis for large fields, indicating that the operator entanglement associated with the $X$- and $Z$-flows behave differently when it comes to targeting the excitation. Our results for Potts are shown in Fig.~(\ref{fig3})(a,b), for $N=50$, in the $Z$-basis, and as compared to TEBD results. Again we see that the approach works remarkably well for large fields, since the $Z$-quasiparticle basis becomes more exact as the field increases.

\begin{figure}
\includegraphics[width=8.5cm,angle=0]{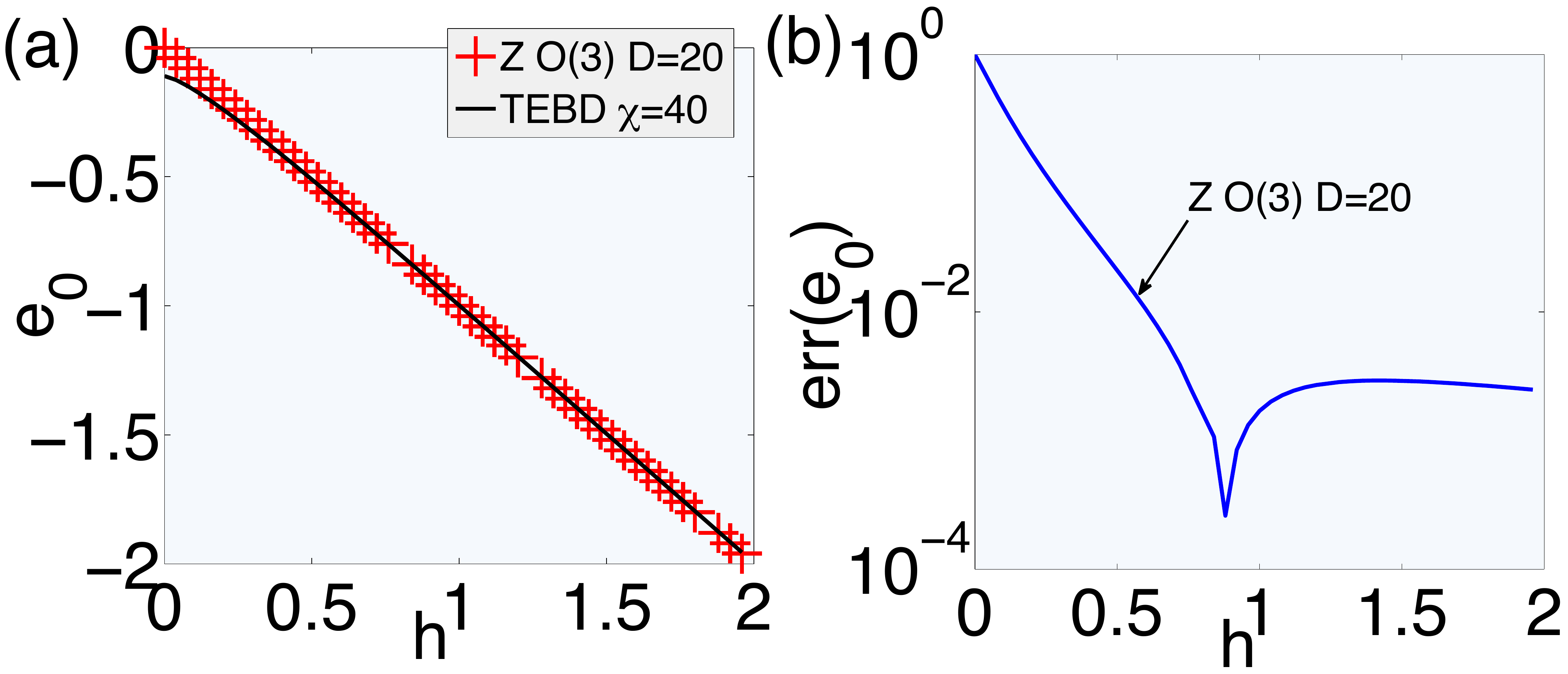}
\caption{[Color online] Quantum 3-state Potts model: (a) ground state energy per site $e_0$ as a function of the magnetic field for $N=50$, all MPO bond dimensions upper-bounded by $D=20$, for simulations in the $Z$  basis and $O(3)$ Taylor order. The critical point in the thermodynamic limit is at $h_{\rm c}=1$. The data is compared with the one obtained by TEBD with MPS bond dimension $\chi=40$ (black line). (b) Relative error in $e_0$ for $N=50$ between eCUT and TEBD for the data in (a).}
\label{fig3}
\end{figure}

\emph{Conclusions and outlook.-} Here we have introduced a scheme to solve the flow equation in CUTs using TNs, based on truncating the operator-entanglement content generated throughout the flow. We have provided proof-of-principle results for 1d systems of size up to $N=50$ for the spin-1/2 quantum Ising and 3-Potts models in a transverse field. Our method allows to extract ground-state energies and, with small modifications, low-energy excitations as well. We believe that this technique has lots of potential. For instance, it is a natural candidate to study excitations in 2d systems \cite{2dtdvp}. This technique will also be fruitful in the study of many-body localized phases \cite{mbl} where, e.g., in 1d, Hamiltonians can be diagonalized by a finite-depth quantum circuit \cite{mblent, pollmbl}. Moreover, our method may be useful in the context of functional-RG as an alternative truncation scheme of the Wetterich RG flow equation \cite{fRG}. Finally, by running the flow backwards one would be able to study how quasiparticles get ``dressed" by quantum correlations in a quantum many-body system. 

\acknowledgements
Discussions with J. Eisert, J. J. Garc\'ia-Ripoll, S. Kehrein, D. G. Olivares, F. Pollmann and M. Rizzi are acknowledged. R. O. and S. S. acknowledge financial support from JGU.

\section{Appendix}

\subsection{Proof of validity for the block generator}

We want to show that the block generator transforms the subspace of the block in the same way as the quasi-particle generator \cite{generators, Knetter2000}. 
To achieve this, we start with the special case $r = 0$, and follow the derivation in Ref.\cite{fischer_description_20112}. We denote the vacuum state with $\ket{0} \equiv (1, 0, \dots, 0)^\intercal$ and, instead of transforming the Hamiltonian
with $\mathcal{H}(\ell) = U(\ell) \mathcal{H}(\ell=0) U^\dagger(\ell)$, we transform the vacuum as $\ket{0(l)} = U(\ell)\ket{0(\ell)}$ (Schr\"odinger picture). The flow of the vacuum state is then given by 
\beq
	\frac{\partial \ket{0(\ell)}}{\partial \ell} =  U(\ell)\underbrace{U^\dagger(\ell)\frac{ \partial U(\ell)}{ \partial \ell}}_{=-\eta(\ell)}\ket{0} = - U(\ell) \eta(\ell) \ket{0}.
\eeq
We can now introduce an orthonormal basis $\{ \ket{n} \}$ and obtain
\beq
	\frac{ \partial \ket{0(\ell)}}{ \partial \ell} = - \sum_n U(\ell) \ket{n}\underbrace{\bra{n} \eta(\ell) \ket{0}}_{\eta_{n0}(\ell)}, 
	\label{eq:flow_state}
\eeq
with the matrix elements $\eta_{n0}$ given by 
\beq
	\eta_{n0}(\ell) = 
	\begin{cases}
		\mathcal{H}_{n0}(\ell) & \text{for } n>0\\
		0 & \text{for } n=0 .
	\end{cases}
\eeq
The above equations yield
\beqa
	\frac{\partial \ket{0(\ell)}}{ \partial \ell} &=& - \left( {\sum_{n} U(\ell) \ket{n}\bra{n} \mathcal{H}(\ell) \ket{0}} \right)  \nonumber \\Ê
	&+& U(\ell) \ket{0}\bra{0} \mathcal{H}(\ell) \ket{0}.
\eeqa
We shift the $\ell$-dependency to the vacuum state and obtain $\frac{\partial \ket{0(\ell)}}{ \partial \ell} = \left[ P_0(\ell), \mathcal{H} \right] \ket{0(\ell)}$, with $\mathcal{H} \equiv \mathcal{H}(\ell=0)$, and where we defined the $\ell$-dependent projector $P_0(\ell) = \ket{0(\ell)}\bra{0(\ell)} $. 

Now we want to generalize the derivation above for $r>0$. We denote the states with $\ket{n}$, e.g, $\ket{0} = (1, 0, \dots, 0)^\intercal$ and $\ket{1} = (0, 1, \dots, 0)^\intercal$. Consider the flow of any state $n < r$, given by 
\beq
	\frac{\partial}{ \partial \ell} \ket{n(\ell)} = -\sum_m U(\ell) \ket{m} \underbrace{\bra{m} \eta(\ell) \ket{n}}_{= \eta_{mn}(\ell)}.
\eeq
The matrix elements $\eta^{\rm qp}_{mn}(\ell)$ for the quasi-particle generator are given by $\eta^{\rm qp}_{mn}(\ell) = \text{sgn}(m - n) \mathcal{H}_{mn}(\ell)$, and for the block generator by
\beq
	\eta^{\rm b}_{mn}(\ell) = 
	\begin{cases}
		\text{sgn}(m - n) \mathcal{H}_{mn}(\ell) &  n \leq r < m; m \leq r < n\\
		0 & n,m \leq r  \text{ and } n,m > r .
	\end{cases}
\eeq
Hence we have for the quasi-particle generator
\beqa
	\frac{\partial}{\partial \ell} \ket{n(\ell)}	&=&- \sum_{m>r}   U(\ell) \ket{m} \bra{m}\mathcal{H}(\ell) \ket{n} \nonumber \\Ê
	&-& \sum_{m \leq r} \text{sgn}(m-n)   U(\ell) \ket{m} \bra{m}\mathcal{H}(\ell)\ket{n}.
\eeqa
By adding and subtracting terms and shifting the $\ell$-dependence to the states one gets
\beqa
	\frac{\partial}{\partial \ell} \ket{n(\ell)} &=& -\mathcal{H} \ket{n(\ell)} + \sum_{m \leq r}  \ket{m(\ell)} \bra{m(\ell)}\mathcal{H}\ket{n(\ell)} \nonumber \\Ê
	&-& \sum_{m \leq r} \text{sgn}(m-n) \ket{m(\ell)} \bra{m(\ell)}\mathcal{H}\ket{n(\ell)}
\label{eq:mielke_state_flow}
\eeqa
For the block-generator we get (after shifting the $\ell$-dependence to the states) 
\beq
	\frac{\partial}{\partial \ell} \ket{n(\ell)} = -\mathcal{H} \ket{n(\ell)} + \sum_{m \leq r}  \ket{m(\ell)} \bra{m(\ell)}\mathcal{H}\ket{n(\ell)}, 
\eeq
which is a generalization of the equation found previously for $r=0$. Notice that the transformation of the subspace with $n \leq r$ for both generators is independent of all the other states with $n > r$, and it only depends on the initial Hamiltonian and the states $n \leq r$ themselves.
The two generators only differ in the last term of Eq.~(\ref{eq:mielke_state_flow}). But this term only includes
the off-diagonal elements in the $(r+1) \times (r+1)$-block, which vanish for the quasi-particle generator but
do not transform for the block generator. Thus both generators transform the considered subspace in the same way, 
but the quasi-particle generator also diagonalizes the block. In our case, we find it convenient to diagonalize exactly the residual small block procuded by the flow. 

\subsection{Details on the numerical Tensor Network approach}

\emph{Truncation.-}  At every step in the algorithm, truncations need to be done in the MPO bond dimension in order to avoid an exponential growth in the number of parameters, which amounts to truncate in the operator-entanglement content. This could be achieved in different ways. For instance, one could do a variational optimization by sweeping throughout the MPO tensors as in DMRG \cite{TN, DMRG}. An alternative strategy, however, is to find the canonical form of the MPO and then truncate in the bond dimension, as in the TEBD algorithm \cite{iTEBD}. In our 1d calculations we have seen that both approaches produce similar results, but the second is a bit more efficient.  

\vspace{5pt}
Ê
\emph{Stability.-}ÊWe have found several sources of instability in the code that need to be very taken into account for large sizes. First, the algorithm is very sensitive to the Taylor order, since errors from the non-unitarity of $U(\ell)$ propagate fast. It is thus important to do a precise truncation in the Taylor series for $U(\ell)$, which in our case is sufficient at order three. The second source of error has to do with the overall norm of the MPOs. In TN algorithms based on the Schr\"odinger's picture, where quantum states are evolved (e.g., TEBD), tensors  can be normalized almost at will throughout the flow in order to keep them numerically well-conditioned, as long as the final observables are computed dividing by the norm of the whole quantum state. However, this is not the situation for eCUT, since it is based on the Heisenberg picture. To be more precise: multiplying the MPO tensors by constants in order to keep them numerically well-conditioned must be done such that \emph{the overall norm of the MPO does not change}, because otherwise one introduces big errors in the Taylor expansion, as well as in the eigenenergies. This turns out to be quite delicate when combined with the canonical form of MPOs mentioned before \cite{iTEBD}, since $\Gamma$-tensors usually have very small components, while $\lambda$-matrices have them very large (and therefore the overall norm of the Hamiltonian MPO is always $O(N)$, with $N$ the size of the system).  In practice, we  found that the following ``tricks" produce well-behaved tensors for large $N$: 
\begin{enumerate} 
\item{After every MPO truncation, normalize $\lambda$s and $\Gamma$s as $\lambda^{[i]} \rightarrow  \lambda^{[i]} / {\rm max} (\lambda^{[i]}) , ~ \Gamma^{[i]} \rightarrow  \Gamma^{[i]} \times {\rm max} (\lambda^{[i]})$, with $i$ a given site. This normalization, which we call \emph{relative normalization to the left}, produces well-conditioned tensors and preserves the overall MPO norm.}
\item{After computing a new MPO, contract $\Gamma$s and $\lambda$s as $A^{[i]}_{\alpha \beta} \equiv \Gamma^{[i]}_{\alpha \beta} \lambda^{[i]}_{\beta \beta}$, and operate with tensors $A^{[i]}$ until the next truncation (i.e., \emph{absorb $\lambda$s to the left}).}
\item{Compute the canonical form by applying until convergence the identity gate to the MPOs \cite{iTEBD}, \emph{sequentially and always from left to right.}}
\item{In the truncation, use an adaptive bond dimension, i.e., remove diagonal values of $\lambda$ below some cutoff $\epsilon$, \emph{being this cutoff proportional to the overall norm of $\lambda$, e.g., $\epsilon = 10^{-15} \times  {\rm max} (\lambda)$}.}
\end{enumerate} 
The combination of the above tricks helps greatly to have well-conditioned tensors throughout the flow while keeping the overall norm of the MPOs constant.  


\begin{thebibliography}{0}

\bibitem{hubbard}Ê
J. Hubbard, Proc. Roy. Soc. (London), Ser. A {\bf 276}, 238 (1963)

\bibitem{highTc}
P. W. Anderson, Science {\bf 235}, 1196 (1987).

\bibitem{KHAF} 
V. Elser, Phys. Rev. Lett. {\bf 62}, 2405 (1989); J. B. Marston and C. Zeng, J. Appl. Phys. {\bf 69}, 5962 (1991); P. Nikolic and T. Senthil, Phys. Rev. B {\bf 68}, 214415 (2003); R. R. P. Singh and D. A. Huse, Phys. Rev. B {\bf 76}, 180407 (2007); ibid, {\bf 77}, 144415 (2008); G. Evenbly and G. Vidal, Phys. Rev. Lett. {\bf 104}, 187203 (2010); S. Yan, D. A. Huse and S. R. White, Science {\bf 332}, 1173-1176 (2011); S. Depenbrock, I. P. McCulloch and U. Schollwoeck, Phys. Rev. Lett. {\bf 109}, 067201 (2012); Z. Y. Xie, J. Chen, J. F. Yu, X. Kong, B. Normand and T. Xiang, Phys. Rev. X {\bf 4}, 011025 (2014); D. Poilblanc and N. Schuch, Phys. Rev. B {\bf 87}, 140407 (2013); D. Poilblanc, N. Schuch, D. P\'erez-Garc\'ia and J. I. Cirac, Phys. Rev. B {\bf 86}, 014404 (2012); T. Picot, M. Ziegler, R. Or\'us and D. Poilblanc, Phys. Rev. B {\bf 93}, 060407 (2016).

\bibitem{sign} 
M. Troyer and U.-Jens Wiese, Phys. Rev. Lett. {\bf 94} 170201 (2005). 

\bibitem{CUT}
F. Wegner, Ann. Phys. (Leipzig) {\bf 3}, 77 (1994); S. D. Glazek and K. G. Wilson, Physical Review D {\bf 49}, 4214 (1994); S. D. Glazek and K. G. Wilson, Physical Review D {\bf 48}, 5863 (1993); S. Kehrein, \emph{The Flow Equation Approach to Many-Particle Systems} (Springer Science \& Business Media, 2006). 

\bibitem{generators}
M. Toda, "Theory of nonlinear lattices", Springer Verlag Berlin (1989); A. Mielke, Eur. Phys. J. B {\bf 5}, 605 (1998); S.R. White, J. Chem. Phys. {\bf 117}, 7472 (2002). 

\bibitem{Knetter2000}
C. Knetter and G. S. Uhrig, Eur. Phys. J. B {\bf 13}, 209 (2000).

\bibitem{deo}
C. M. Dawson, J. Eisert, and T. J. Osborne, Phys. Rev. Lett. {\bf 100}, 130501 (2008). 

\bibitem{pcut} 
J. Stein, J. Stat. Phys. {\bf 88}, 487 (1997); C. Knetter, K. P. Schmidt and G. S. Uhrig, J. Phys. A {\bf 36}, 7889 (2003); A. Hackl and S. Kehrein, Phys. Rev. B {\bf 78}, 092303 (2008); H. Krull, N.A. Drescher, and G.S. Uhrig, Phys. Rev. B {\bf 86}, 125113 (2012).

\bibitem{gcut}
H.-Y. Yang and K. P. Schmidt, Eur. Phys. Lett. {\bf 94}, 17004 (2011); K. Coester, S. Clever, F. Herbst, S. Capponi, and K. P. Schmidt, Eur. Phys. Lett. {\bf 110}, 20006 (2015). 

\bibitem{sCUT}
T. Fischer, S. Duffe, and G.S. Uhrig, New J. Phys. {\bf 12}, 033048 (2010); N.A. Drescher, T. Fischer, and G.S. Uhrig, Eur. Phys. J. B {\bf 79}, 225 (2011).

\bibitem{Heidbrink2002}
C.P. Heidbrink and G.S. Uhrig, Phys. Rev. Lett. {\bf 88}, 14601 (2002).

\bibitem{scalingdim}
S. Kehrein, Phys. Rev. Lett. {\bf 83}, 4914 (1999); S. Kehrein, Nucl. Phys. B {\bf 592}, 512 (2001); M. Powalski, G.S. Uhrig, and K.P. Schmidt, Phys. Rev. Lett. {\bf 115}, 207202 (2015).

\bibitem{TN}
R. Or\'us, Annals of Physics {\bf 349} 117-158 (2014); J. Eisert, Modelling and Sim. {\bf 3}, 520 (2013); N. Schuch, QIP, Lecture Notes of the 44th IFF Spring School 2013; J. I Cirac and F. Verstraete, J. Phys. A: Math. Theor. {\bf 42}, 504004 (2009); F. Verstraete, J. I. Cirac and V. Murg, Adv. Phys. {\bf 57}, 143 (2008).

\bibitem{DMRG}
S. R. White, Phys. Rev. Lett. {\bf 69}, 2863 (1992); S. R. White, Phys. Rev. B {\bf 48}, 10345 (1993). 

\bibitem{tebd}
G. Vidal, Phys. Rev. Lett. {\bf 91}, 147902 (2003); G. Vidal, Phys. Rev. Lett. {\bf 93}, 040502 (2004). 

\bibitem{PEPS}
F. Verstraete and J. I. Cirac, cond-mat/0407066; V. Murg, F. Verstraete and J. I. Cirac, Phys. Rev. A {\bf 75}, 033605 (2007). 

\bibitem{MERA}
G. Vidal, Phys. Rev. Lett. {\bf 99}, 220405 (2007); For an introduction, see, e.g., G. Vidal, chapter of the book \emph{Understanding Quantum Phase Transitions}, edited by Lincoln D. Carr (Taylor \& Francis, Boca Raton, 2010), arXiv:0912.1651v2.

\bibitem{iTEBD}
G. Vidal, Phys. Rev. Lett. {\bf 91}, 147902 (2003); G. Vidal, Phys. Rev. Lett. {\bf 93}, 040502 (2004); G. Vidal, Phys. Rev. Lett. {\bf 98}, 070201 (2007); R. Or\'us and G. Vidal, Phys. Rev. B {\bf 78}, 155117 (2008). 

\bibitem{iPEPS}
J. Jordan, R. Or\'us, G. Vidal, F. Verstraete, and J. I. Cirac, Phys. Rev. Lett. {\bf 101}, 250602 (2008); R. Or\'us and G. Vidal, Phys. Rev. B {\bf 80}, 094403 (2009); H. N. Phien, J. A. Bengua, H. D. Tuan, P. Corboz, and R. Or\'us, Phys. Rev. B {\bf 92}, 035142 (2015). 

\bibitem{iDMRG}
I. P. McCulloch, arXiv:0804.2509; G. M. Crosswhite, A. C. Doherty and G. Vidal, Phys. Rev. B {\bf 78}, 035116 (2008).

\bibitem{fermions}
P. Corboz, G. Evenbly, F. Verstraete and G. Vidal, Phys. Rev. A {\bf 81}, 010303(R) (2010); P. Corboz and G. Vidal, Phys. Rev. B {\bf 80}, 165129 (2009); P. Corboz, R. Or\'us, B. Bauer and G. Vidal, Phys. Rev. B {\bf 81}, 165104 (2010); C. V. Kraus, N. Schuch, F. Verstraete and J. I. Cirac, Phys. Rev. A {\bf 81}, 052338 (2010); I. Pizorn and F. Verstraete, Phys. Rev. B {\bf 81}, 245110 (2010); Q.-Q. Shi, S.-Hao Li, J.-Hui Zhao and H.- Qiang Zhou, arXiv:0907.5520; C. Pineda, T. Barthel and J. Eisert, Phys. Rev. A {\bf 81}, 050303(R) (2010); T. Barthel, C. Pineda and J. Eisert, Phys. Rev. A {\bf 80}, 042333 (2009). 

\bibitem{mps}
F. Verstraete and J. I. Cirac, Phys. Rev. B {\bf 73}, 094423 (2006). 

\bibitem{flowMERA}
C. M. Dawson, J. Eisert, and T. J. Osborne, Phys. Rev. Lett. {\bf 100}, 130501 (2008).

\bibitem{cMERA} 
J. Haegeman, T. J. Osborne, H. Verschelde, and F. Verstrate, Phys. Rev. Lett. {\bf 110}, 100402 (2013). 

\bibitem{2dtdvp}
L. Vanderstraeten, M. Mari\"en, F. Verstraete and J. Haegeman, Phys. Rev. B {\bf 92}, 201111 (2015).

\bibitem{mbl}
R. Nandkishore, D. H. Huse, Annual Review of Condensed Matter Physics, {\bf 6}: 15-38 (2015); C. Monthus, 
J. Phys. A: Math. Theor. {\bf 49} 305002 (2016). 

\bibitem{mblent}
B. Bauer and C. Nayak, J. Stat. Mech. (2013) P09005. 

\bibitem{pollmbl}
 F. Pollmann, V. Khemani, J. I. Cirac and S. L. Sondhi, Phys. Rev. B {\bf 94}, 041116 (2016). 
 
\bibitem{fRG}
J. Berges, N. Tetradis, and C. Wetterich, Phys. Rept. {\bf 363}:223-386 (2002). 

\bibitem{fischer_description_20112}
T. Fischer, \emph{Description of quasiparticle decay by continuous unitary transformations}, PhD thesis, Technische Universit\"at Dortmund, 2011.


\end{thebibliography}
\end{document}